# Thermosize potentials in semiconductors

S. Karabetoglu, A. Sisman*
Nano Energy Research Group
Istanbul Technical University, Energy Institute, 34469, Istanbul/Turkey
*Corresponding Author: sismanal@itu.edu.tr

**Abstract**
A thermosize junction consists of two different sized structures made by the same material. Classical and quantum thermosize effects (CTSE and QTSE), which are opposite to each other, induce a thermosize potential in a thermosize junction. A semi-analytical method is proposed to calculate thermosize potentials in wide ranges of degeneracy and confinement by considering both CTSE and QTSE in thermosize junctions made by semiconductors. Dependencies of thermosize potential on temperature, size and degeneracy are examined. It is shown that a potential difference in millivolt scale can be induced as a combined effect of CTSE and QTSE. The highest potential is obtained in non-degenerate limit where the full analytical solution is obtained. The model can be used to design semiconductor thermosize devices for a possible experimental verification of CTSE and QTSE, which may lead to new nano energy conversion devices.

## 1. Introduction

An electrical potential difference is induced by a temperature gradient in a junction made by different types of semiconductors or conductors. This is a well-known thermoelectric Seebeck effect and it is used to convert heat directly into electricity in thermoelectric generators. Similarly, thermosize potential is induced in a junction made by the same kind of material but having different sizes under the same temperature gradient. Thermosize effects (TSE) are the analogue of thermoelectric effects and they have been proposed by Sisman and Muller as a consequence of quantum size effects (QSE) resulting from the wave character of particles [1]. After the proposition of quantum based TSE, also classical TSE have been proposed by Babac and Sisman as a result of different transport regimes [2] in macro and nano scaled structures. Later, TSE have been classified into two groups as classical and quantum TSE, (CTSE and QTSE) [3].

Because of classical and quantum size effects, size itself becomes a control parameter on transport properties of a material, especially at nanoscale. Therefore the same material has different properties for its different sizes. One of the most important and well-known classical size effects is due to the change of transport regime from hydrodynamic to ballistic when the characteristic size of a material is getting smaller in comparison with the mean free path of particles, $l$. Even for electrons in semiconductors, it is possible to use hydrodynamic description when electron-electron mean free path $l_{ee}$ is smaller than mean free paths for electron-impurity $l_{ei}$ and electron-phonon $l_{ep}$ collisions as well as the material's characteristic size [4]. CTSE appear due to difference in transport regimes when two different sized structures made by the same material are bring together to form a junction under a temperature gradient. CTSE reach to their maximum strength if one of the structures has ballistic regime while the other has hydrodynamic one. Similarly, QSE become noticeable and make thermodynamic and transport properties size and shape dependent when at least one of the sizes of a material is in the order of thermal de Broglie wavelength of particles, $\lambda$, [1, 3, 5-15]. QTSE appear due to difference in QSE when two different sized structures made by the same material are under a temperature gradient. They also reach to their maximum strength if one of the domain is strongly confined at least in one direction while the other is free in all directions.

In most cases, it is difficult to observe pure QTSE without CTSE since the mean free path of carriers is usually in the same order of magnitude with their thermal de Broglie wavelength in semiconductors [16-19]. Therefore CTSE and QTSE are usually coexist and unfortunately they induce thermosize potentials opposite to each other. Consequently, it is important to develop a model to consider both effects and then isolate one from another mathematically for their possible experimental verifications. In recent years, there are some studies on thermodynamic analysis of conceptual thermosize cycles as well as some new cycles based on QTSE and CTSE [20-29]. All these studies consider either CTSE or QTSE alone although they coexist in general and the derivations are based on only weakly confined ideal atomic gases. However, the most practical way for a possible experimental verification of TSE is to measure the total (CTSE and QTSE) thermosize potential in strongly confined structures made by semiconductors or conductors instead of trying to indirect measurement of the chemical potential difference due to CTSE or QTSE alone in weakly confined atomic gases. Therefore, calculation of total thermosize potential in strongly confined and degenerate Fermi gases is an important issue. In this study, a semi-analytical method is proposed to calculate the total thermosize potential by numerical solution of a derived differential equation which considers both quantum and classical size effects for strongly confined and degenerate ideal Fermi gases. Total thermosize potential is calculated for junctions made by combination of 3D semiconductor structures with 1D or 2D ones. Classical and quantum parts of total potential are decomposed from each other and the variations of their contributions with degeneracy, confinement and temperature are examined.

A single thermosize junction consists of two different sized structures of a single material. A schematic view of a simple thermosize junction under a temperature gradient is shown in Fig.1. In this work, nano domain is considered as 2D (1D) by assuming a strong confinement(s) in one (both) of the transverse directions which are perpendicular to temperature gradient. However, macro domain is considered as a bulk material (3D) by completely neglecting the confinement effects in all directions.

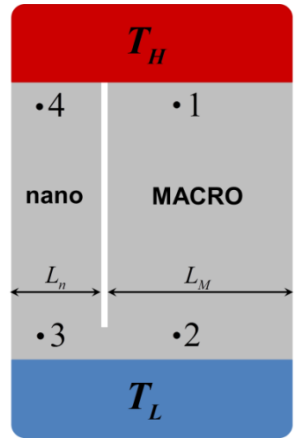

**Fig.1.** A schematic view of a thermosize junction.

An electrochemical potential difference occurs between points 4 and 1 since nano domain is under both classical and quantum size effects while macro one is free of size effects. Therefore the same temperature gradient induces different density gradients in each part. At nano part, ballistic transport regime under quantum size effects determines the value of electrochemical potential (or density) gradient induced by the temperature gradient. At macro part, however, hydrodynamic transport regime under the lack of size effects is responsible for that. This difference of electrochemical potential gradients in nano and macro parts causes a net electrochemical potential difference between point 4 and point 1 since the potentials at point 3 and point 2 are equal to each other due to the connection in between them. This net potential difference is called thermosize potential.

## 2. Model

For the derivations, a rectangular domain geometry is chosen here since it is appropriate geometry for nano manufacturing, besides, it allows to obtain semi-analytical results. Charged particle flux can be controlled by both temperature and electrochemical potential gradients. When the net particle flux is zero, temperature gradient is compensated by a reverse electrochemical potential gradient. Therefore, if a temperature gradient is applied to a domain, it is possible to calculate electrochemical potential distribution in terms of temperature distribution in the domain where the net particle flux is zero. The following differential equation for temperature dependency of electrochemical potential can be derived for zero net particle flux by using the particle flux expression given in Ref. [14], which has been obtained under the relaxation time approximation by considering QSE,

$$\frac{\partial \mu}{\partial T} - \frac{\mu}{T} = -k_b \frac{g_2}{g_0}, \qquad (1)$$

where $\mu$ is electrochemical potential, $T$ is temperature, $k_b$ is Boltzmann's constant and $g$ is a dimensionless function given by,

$$g_a = \sum_{\{i_1,i_2,i_3\}=1}^{\infty} \frac{\gamma_s f_0'(1-f_0')(\alpha_1 i_1)^2}{\left[(\alpha_1 i_1)^2 + (\alpha_2 i_2)^2 + (\alpha_3 i_3)^2\right]^{\frac{1-a}{2}}} \qquad (2)$$

where $f_0'$ is the equilibrium distribution function per spin states and it is given for Fermi statistics as,

$$f_0' = \frac{f_0}{\gamma_s} = \frac{1}{\exp\left[-\Lambda + (\alpha_1 i_1)^2 + (\alpha_2 i_2)^2 + (\alpha_3 i_3)^2\right]+1}, \qquad (3)$$

where $\gamma_s$ is the spin degree of freedom for carriers ($\gamma_s = 2$ for electrons and holes) and $\Lambda = \mu/k_b T$ is dimensionless electrochemical potential. It should be noted that the direction 1 is chosen as the direction of temperature gradient. Since the considered domain has a rectangular shape, dimensionless form of translational kinetic energy of a carrier, $\tilde{\varepsilon}$, is obtained from the trivial solution of Schrödinger equation and represented by

$$\tilde{\varepsilon} = \frac{\varepsilon_{i_1 i_2 i_3}}{k_b T} = (\alpha_1 i_1)^2 + (\alpha_2 i_2)^2 + (\alpha_3 i_3)^2, \quad \{i_1,i_2,i_3\} = 1,2,3... \qquad (4)$$

where $\alpha_1 = L_c/L_1$, $\alpha_2 = L_c/L_2$, $\alpha_3 = L_c/L_3$ are confinement parameters and $L_c = h/2\sqrt{2m^* k_b T}$ is a length scale based on de Broglie wavelength of carriers with effective mass of $m^*$. By applying the first two terms of Poisson summation formula for calculation of the summations in Eq.(2), the following analytical expression for the ratio of $g$ functions is obtained,

$$\frac{g_2}{g_0} = \frac{2Li_2\left[-\exp(\Lambda)\right]}{Li_1\left[-\exp(\Lambda)\right]} \left[\frac{1 - \frac{9\sqrt{\pi}(\alpha_2+\alpha_3)}{32}\frac{Li_{3/2}\left[-\exp(\Lambda)\right]}{Li_2\left[-\exp(\Lambda)\right]} + \frac{3(\alpha_2\alpha_3)}{4\pi}\frac{Li_1\left[-\exp(\Lambda)\right]}{Li_2\left[-\exp(\Lambda)\right]}}{1 - \frac{3\sqrt{\pi}(\alpha_2+\alpha_3)}{8}\frac{Li_{1/2}\left[-\exp(\Lambda)\right]}{Li_1\left[-\exp(\Lambda)\right]} + \frac{3(\alpha_2\alpha_3)}{2\pi}\frac{Li_0\left[-\exp(\Lambda)\right]}{Li_1\left[-\exp(\Lambda)\right]}}\right], \qquad (5)$$

where $Li_n\left[-\exp(\Lambda)\right]$ is the Polylogarithm function with the degree of $n$ and argument of $-\exp(\Lambda)$. It should be noted that confinement parameters, $\alpha$, are inversely proportional to the square root of temperature and $g_2/g_0$ is a complicated function of electrochemical potential as it is seen in Eq.(5). Numerical calculations of complete forms of $g$ functions showed that the relative errors of Eq.(5) are 5.5% and 0.1% when $\Lambda = -5$ and $\Lambda = 10$ respectively for 2D case, $\alpha_3 = 1$ and $\alpha_2 = \alpha_1 = 0.01$. The errors become 8.4% and 0.1% respectively for the same values of chemical potentials in case of 1D, $\alpha_3 = \alpha_2 = 1$ and $\alpha_1 = 0.01$. It is clear that these errors become even smaller when confinement parameters become smaller. Therefore, instead of using full numerical solution based on Eqs.(1) and (2), the semi-analytical solution based on Eq.(5) can reliably be used since it provides much faster calculations in comparison with the full numerical one. Furthermore, semi-analytical solution gives a physical insight and allows to make analytical interpretations of the results since functional dependencies on confinement parameters and chemical potential are explicitly known. By using Eq.(5) in Eq.(1), the following differential equation for temperature and confinement dependencies of $\Lambda$ can be written

$$\frac{d\Lambda}{dT} = -\frac{1}{T}\frac{2Li_2[-\exp(\Lambda)]}{Li_1[-\exp(\Lambda)]}\left[\frac{1-\frac{9\sqrt{\pi}(\alpha_2+\alpha_3)}{32}\frac{Li_{3/2}[-\exp(\Lambda)]}{Li_2[-\exp(\Lambda)]}+\frac{3(\alpha_2\alpha_3)}{4\pi}\frac{Li_1[-\exp(\Lambda)]}{Li_2[-\exp(\Lambda)]}}{1-\frac{3\sqrt{\pi}(\alpha_2+\alpha_3)}{8}\frac{Li_{1/2}[-\exp(\Lambda)]}{Li_1[-\exp(\Lambda)]}+\frac{3(\alpha_2\alpha_3)}{2\pi}\frac{Li_0[-\exp(\Lambda)]}{Li_1[-\exp(\Lambda)]}}\right]. \tag{6}$$

with the boundary condition of $\Lambda = \Lambda_L$ at $T = T_L$. Therefore, temperature and confinement dependencies of electrochemical potential for nano part of the domain can be determined only by numerical solution of Eq.(6) and then the solution can be represented in the following implicit form,

$$T^2 Li_2[-\exp(\Lambda)][1+h(\Lambda,\alpha_2,\alpha_3)] = \text{const}. \tag{7}$$

where the function $h(\Lambda,\alpha_2,\alpha_3)$ stands for the implicit form of the correction due to QSE and it can only be numerically determined. It vanishes when all confinement parameters go to zero, $\{\alpha_2,\alpha_3\} \to 0$. In case of confinement parameters are negligible, Eq.(6) becomes a simple expression and analytical solution of Eq.(6) becomes possible. In this case; the generalized Knudsen law, which is valid for both quantum and classical gases, is obtained from the solution of Eq.(6) and it is directly recovered from Eq.(7) just by neglecting $h(\Lambda,\alpha_2,\alpha_3)$, [2]. Therefore the function $h(\Lambda,\alpha_2,\alpha_3)$ represents the deviations from the generalized Knudsen law due to QSE.

In macro part, all confinement parameters go to zero and also the transport regime is defined as hydrodynamic regime because the characteristic length of the domain is much bigger than the mean free path of charge carriers. When the net particle flux is zero, pressure becomes a constant quantity in case of hydrodynamic regime and the relation between temperature and electrochemical potential is simply given by [2],

$$T^{5/2} Li_{5/2}[-\exp(\Lambda)] = \text{const}. \tag{8}$$

By using temperature and electrochemical potentials of points 3 and 4 in Eq.(7) and those of points 1 and 2 in Eq.(8), the following equations are obtained respectively,

$$T_H^2 Li_2[-\exp(\Lambda_4)][1+h(\Lambda_4,\alpha_2,\alpha_3)] = T_L^2 Li_2[-\exp(\Lambda_3)][1+h(\Lambda_3,\alpha_2,\alpha_3)] \tag{9a}$$

$$T_H^{5/2} Li_{5/2}[-\exp(\Lambda_1)] = T_L^{5/2} Li_{5/2}[-\exp(\Lambda_2)] \tag{9b}$$

Because of the connection at low temperature side of thermosize junction, electrochemical potentials are equal to each other at low temperature part, $\Lambda_2 = \Lambda_3 = \Lambda_L$. Therefore it is possible to calculate $\Lambda_4$ in terms of $T_L$, $T_H$, $\Lambda_L$, $\alpha_2$ and $\alpha_3$ by numerical solution of Eqs.(9a). Similarly, $\Lambda_1$ can be calculated as functions of $T_L$, $T_H$ and $\Lambda_L$ by solution of Eq.(9b). Consequently, the induced thermosize potential at high temperature side is determined by just subtracting electrochemical potentials of points 4 and 1 from each other as,

$$\Delta\mu_{TSE} = k_b T_H (\Lambda_4 - \Lambda_1) \tag{10}$$

Furthermore, thermosize potential can be split into two parts as contributions of CTSE and QTSE. Therefore Eq.(10) can be written as a summation of two following equations, Eqs.(11) and (12), which represent quantum and classical thermosize potentials respectively,

$$\Delta\mu_{QTSE} = k_b T_H (\Lambda_4 - \Lambda_4^o) \tag{11}$$

$$\Delta\mu_{CTSE} = k_b T_H \left( \Lambda_4^o - \Lambda_1 \right) \tag{12}$$

where $\Lambda_4^o$ is the electrochemical potential of point 4 in case of confinement parameters are neglected. In other words, it is calculated by considering only CTSE from the following equation

$$T_H^2 Li_2\left[-\exp(\Lambda_4^o)\right] = T_L^2 Li_2\left[-\exp(\Lambda_L)\right] \tag{13}$$

In case of non-degeneracy, Eq.(6) reduces a simpler differential equation since all Polylogarithm functions become a simple expression, $Li_n\left[-\exp(\Lambda)\right] \to -\exp(\Lambda)$, and cancel each other in asymptotic limit, $\Lambda \to -\infty$. The reduced form of Eq.(6) can analytically be solved and electrochemical potential difference for nano part is given as

$$\Lambda_4 - \Lambda_L = 2\ln\left(\frac{T_L}{T_H}\right) + \ln\left[\frac{1 - \frac{3\sqrt{\pi}\left(\alpha_2^L + \alpha_3^L\right)}{8} + \frac{3\alpha_2^L \alpha_3^L}{2\pi}}{1 - \frac{3\sqrt{\pi}\left(\alpha_2^H + \alpha_3^H\right)}{8} + \frac{3\alpha_2^H \alpha_3^H}{2\pi}}\right] \tag{14}$$

Similarly, by using the asymptotic forms of Polylogarithm functions, electrochemical potential difference for macro part is obtained from Eq.(9b) as

$$\Lambda_1 - \Lambda_L = \frac{5}{2}\ln\left(\frac{T_L}{T_H}\right). \tag{15}$$

Therefore, use of Eqs.(14) and (15) in Eq.(10) simply gives

$$\Delta\mu_{TSE} = -\frac{1}{2} k_b T_H \ln\left(\frac{T_L}{T_H}\right) + k_b T_H \ln\left[\frac{1 - \frac{3\sqrt{\pi}\left(\alpha_2^L + \alpha_3^L\right)}{8} + \frac{3\alpha_2^L \alpha_3^L}{2\pi}}{1 - \frac{3\sqrt{\pi}\left(\alpha_2^H + \alpha_3^H\right)}{8} + \frac{3\alpha_2^H \alpha_3^H}{2\pi}}\right] \tag{16}$$

The first and second terms of the right hand side of Eq.(16) represent $\Delta\mu_{CTSE}$ and $\Delta\mu_{QTSE}$ respectively in non-degenerate limit. For weakly confined 2D case ($\alpha_2 \ll 1$ and $\alpha_3 = 0$), after some mathematical manipulations, Eq.(16) reduces to the know expression in literature [3]. Therefore, for non-degenerate case, Eq.(16) represents the full analytical solution.

## 3. Results and Discussion

Variations of thermosize potentials, represented by Eqs.(10)-(12), with cold side degeneracy are shown in Fig. 2 for 2D/3D and 1D/3D confinement configurations. The confinement parameters of 2D nano domain are chosen as $\alpha_3^L = 1$, (strongly confined at cold side in one of the transverse directions) and $\alpha_1 = \alpha_2 = 0.01$ (nearly free in other directions) while they are $\alpha_1 = \alpha_2 = \alpha_3 = 0.01$ for macro domain. For 1D nano domain, strong confinement in both transverse directions of cold side are considered, $\alpha_3^L = \alpha_2^L = 1$. For both non-degenerate and degenerate cases, it is seen that quantum thermosize potential has an opposite sign of classical thermosize potential. Therefore, total thermosize potential is always smaller than either its classical or quantum components and it has negative values for non-degenerate case while the situation is reversed for highly degenerate case. Since de Broglie wavelength of charge carriers becomes shorter for higher degeneracy, both QTSE and CTSE become weaker although QTSE is effected more strongly. This makes CTSE dominant and total thermosize potential positive in case of high degeneracy. It is seen that there is a critical value of cold side chemical potential for which CTSE and QTSE cancel each other and induce net zero potential. This turning point of total thermosize potential can be a good indicator for experimental verification of CTSE and QTSE. Furthermore, the potentials induced by QTSE in 1D case are always higher than those in 2D case. For weakly confined case, $\{\alpha_3, \alpha_2, \alpha_1\} \ll 1$, it is clear that QTSE become completely negligible and thermosize potential essentially consists of only CTSE as expected. It seems that induced potentials for all cases are in a measurable scale.

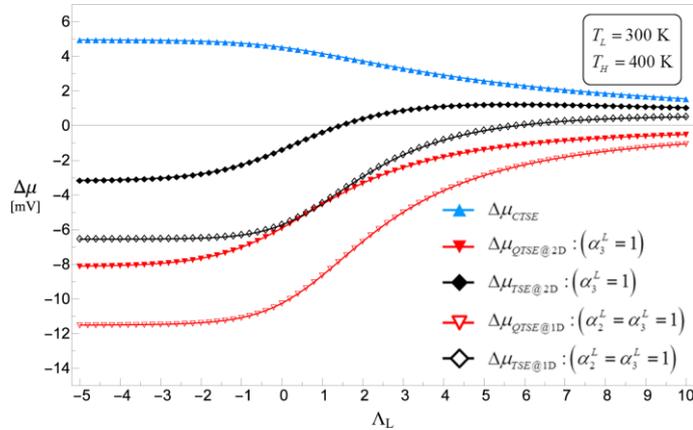

**Fig. 2.** Degeneracy dependency of thermosize potentials for 2D/3D and 1D/3D configurations.

Temperature dependencies of thermosize potentials are shown in Fig. 3 and Fig. 4 for non-degenerate and degenerate cases respectively. Higher temperatures lead to lower confinement at hot side and the factor of $T_H$ in Eq.(16) causes almost linear behaviors of both CTSE and QTSE for non-degenerate case. The similar linear behavior is seen also for degenerate case although the smaller potentials are induced, Fig. 4. It should be noted that total thermosize potential is in a measurable scale if such a structure is made and exposed to a reasonable temperature gradient.

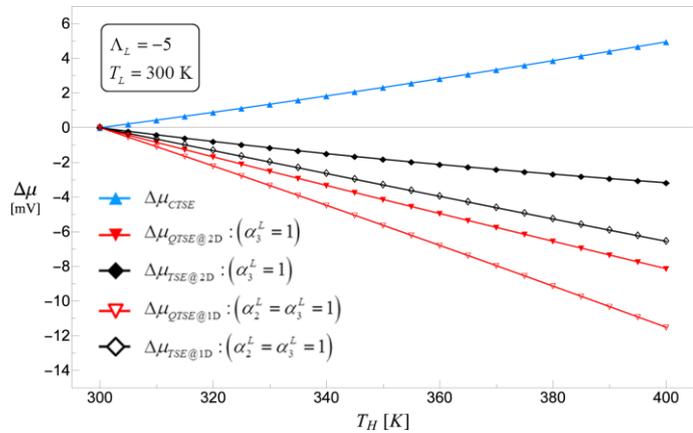

**Fig. 3.** Temperature dependency of thermosize potentials for non-degenerate 2D and 1D cases.

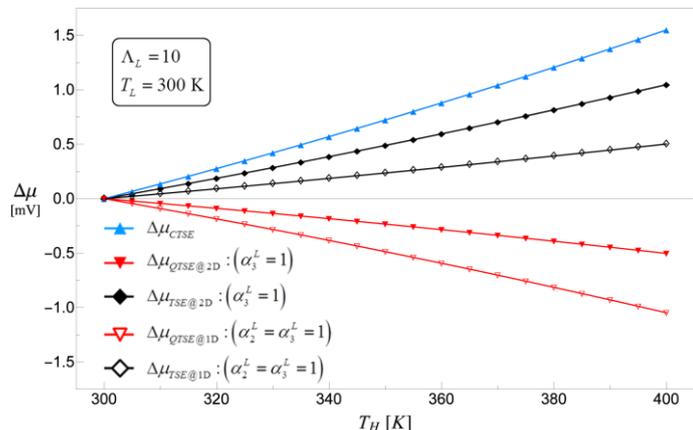

**Fig. 4.** Temperature dependency of thermosize potentials for degenerate 2D and 1D cases

In case of 300 K and 400 K temperatures for cold and hot sides, Fig. 5 shows the variations of thermosize potentials with the value of transverse confinement coefficient of cold side for a specific degeneracy of $\Lambda_L = 0.14$. This value of degeneracy corresponds to maximum difference between $\Delta\mu_{QTSE@1D}$ and $\Delta\mu_{QTSE@2D}$ in Fig. 2. It is clearly seen that CTSE has a constant value while magnitude of QTSE increases for stronger confinement conditions. Total thermosize potential changes its sign during the transition from 3D to 2D and there is a specific confinement value for which total potential becomes zero. It is a turning point of total thermosize potential. Therefore, for a possible experimental verification, this special behavior can be a good indicator for the combined effect of CTSE and QTSE.

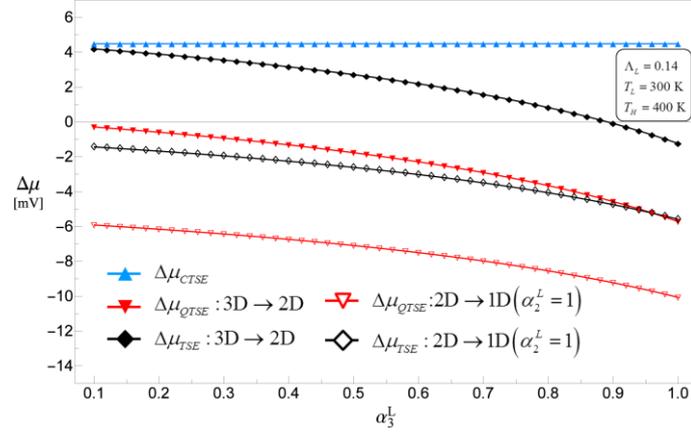

**Fig. 5.** Dependency of thermosize potential on cold side confinements in transverse directions for slightly degenerate case

## 4. Conclusion

Within the cases considered here, strongly confined non-degenerate case promises the highest thermosize potential. Although QTSE induce a higher thermosize potential for strongly confined non-degenerate case, the net termosize potential is smaller than QTSE since CTSE also induce a high potential in reverse direction. On the other hand, if the transverse confinement parameter takes the values much higher than unity, then a very high thermosize potential can be induced since CTSE remains constant while QTSE increases with increasing confinement. For a possible experimental verification, various thermosize junctions made by semiconductors having different transverse confinements can be exposed to the same temperature gradient and size dependency of thermosize potential can be measured to investigate whether the functional behavior is similar to one given in Fig. 5. Another experimental investigation of CTSE and QTSE can be to verify the turning point of total thermosize potential in Fig. 2.

In case of $T_L = 300\,\text{K}$ and $T_H = 400\,\text{K}$, calculated thermosize potentials are given in Table 1 for two different degeneracy values and junction configurations (2D/3D and 1D/3D). The induced thermosize potentials per unit junction thickness ($\delta$) is symbolized by $\Delta\mu'$. Thickness is chosen as 300 nm as a summation of 3.8 nm and 296.2 nm for nano and macro parts respectively. Since electron-electron mean free path is mostly in the order of around 30 nm, 296.2 nm thickness ensures the validity of hydrodynamic transport regime in macro part. Similarly 3.8 nm thickness is enough to make the confinement parameters equal to unity at 300 K. It is seen that thermosize potentials in non-degenerate semiconductor junctions are always higher since degeneracy reduces both QTSE and CTSE. Furthermore, the highest thermosize potential is induced for non-degenerate condition by 1D/3D junction configuration. It should be noted that effective mass of charge carriers depend on material composition of semiconductors and a bare mass of electron is considered during the calculations of the quantities given in Table 1 to make material independent comparison of the values. On the other hand, the results in the figures are independent from effective mass since they depend on confinement parameters itself instead of effective mass and material thickness individually. For a possible experimental verification, the effective masses of charge carriers for the chosen semiconductor have to be considered during the calculations of the values in Table 1.

**Table 1**
Thermosize potentials per unit junction thickness for two different configurations (2D/3D and 1D/3D) and degeneracy conditions in case of $T_L = 300\,\text{K}$ and $T_H = 400\,\text{K}$.

| $\Lambda_L$ | 2D/3D ($L_n = 3.8$ nm  $L_M = 296.2$ nm) | | | 1D/3D ($L_n \times L_n = 3.8 \times 3.8$ nm  $L_M = 296.2$ nm) | | |
|---|---|---|---|---|---|---|
| | $\Delta\mu'_{TSE}$ [μV/nm] | $\Delta\mu'_{CTSE}$ [μV/nm] | $\Delta\mu'_{QTSE}$ [μV/nm] | $\Delta\mu'_{TSE}$ [μV/nm] | $\Delta\mu'_{CTSE}$ [μV/nm] | $\Delta\mu'_{QTSE}$ [μV/nm] |
| −5 | −10.5 | 16.5 | −27.0 | −21.8 | 16.5 | −38.3 |
| 10 | 3.5 | 5.2 | −1.7 | 1.7 | 5.2 | −3.5 |

The proposed model can be used to design and produce prototypes of semiconductor thermosize devices for a possible experimental verification of CTSE and QTSE. A possible verification of thermosize potential will not only prove CTSE and QTSE but also be a demonstration of macroscopic manifestation of quantum size effects. After experimental verification, nano energy conversion devices based on thermosize effects can be designed and produced.